\documentclass[10pt]{article}

\usepackage{authblk}
\usepackage[square,numbers]{natbib}
\usepackage[T1]{fontenc}
\usepackage[utf8]{inputenc}
\usepackage{amsmath,amssymb,amsfonts}
\usepackage{algorithm}
\usepackage{algpseudocode}
\usepackage{graphicx}
\usepackage[caption=false,font=footnotesize]{subfig}
\usepackage{caption}
\usepackage{textcomp}
\usepackage{url}
\usepackage{xurl}
\usepackage[htt]{hyphenat}
\usepackage{xcolor}
\usepackage{array}
\usepackage{booktabs}
\usepackage{braket}
\usepackage{enumitem}
\usepackage{ifthen}
\usepackage{makecell}
\usepackage{multirow}
\usepackage{pifont}
\usepackage{ragged2e}
\usepackage{tcolorbox}
\usepackage{xspace}
\usepackage{setspace}
\usepackage[hidelinks]{hyperref}
\usepackage{geometry}

\newcommand{\Description}[1]{}

\newcolumntype{C}[1]{>{\centering\arraybackslash}p{#1}}

\usepackage{amsmath}
\usepackage{algorithm}
\usepackage{algpseudocode}
\usepackage{graphicx}
\usepackage[caption=false,font=footnotesize]{subfig}
\usepackage{textcomp}
\usepackage{url}
\usepackage{xcolor}
\usepackage{array}
\usepackage{booktabs}
\usepackage{braket}
\usepackage{enumitem}
\usepackage{ifthen}
\usepackage{makecell}
\usepackage{multirow}
\usepackage{pifont}
\usepackage{ragged2e}
\usepackage{tcolorbox}
\usepackage{xspace}
\usepackage{setspace}
\hypersetup{hidelinks}

\def\BibTeX{{\rm B\kern-.05em{\sc i\kern-.025em b}\kern-.08em
    T\kern-.1667em\lower.7ex\hbox{E}\kern-.125emX}}

\definecolor{light-gray}{gray}{0.80}

\newcommand{\AnswerToRQ}[2]{%
  
  \begin{tcolorbox}[
    colback=black!1!white,
    colframe=black!40!white,
    left=0.25mm,
    right=0.25mm,
    top=0.25mm,
    bottom=0.25mm,
    boxsep=0.66mm,
    arc=0.1mm,
  ]
  \textbf{Answer to RQ#1:} #2
  \end{tcolorbox}
  
}

\newcommand{\ApproachName}{\textsf{Qomino}\xspace}
\newcommand{\ChiSquare}{\textsf{Chi2}\xspace}
\newcommand{\JSDiv}{\textsf{JSDiv}\xspace}
\newcommand{\QOPS}{\textsf{QOPS}\xspace}
\newcommand{\DOSS}{\textsf{DOSS}\xspace}
\newcommand{\Inverse}{\textsf{Inverse}\xspace}
\newcommand{\QOIN}{\textsf{QOIN}\xspace}
\newcommand{\Quito}{\textsf{Quito}\xspace}
\newcommand{\QuSBT}{\textsf{QuSBT}\xspace}
\newcommand{\QuCAT}{\textsf{QuCAT}\xspace}
\newcommand{\VariantName}[1]{\texttt{#1}\xspace}
\newcommand{\NoBipartition}{\VariantName{w/o bipart}}
\newcommand{\NoDecomposition}{\VariantName{w/o decom}}
\newcommand{\NoDominance}{\VariantName{w/o domin}}
\newcommand{\NoHoeffding}{\VariantName{w/o hoeff}}
\newcommand{\InconclusiveToPass}{\VariantName{incon2pass}}

\newcommand{\LPR}{\texttt{LPR}\xspace}
\newcommand{\LAF}{\texttt{LAF}\xspace}
\newcommand{\QFT}{\texttt{QFT}\xspace}
\newcommand{\DO}{\texttt{DO}\xspace}
\newcommand{\IC}{\texttt{IC}\xspace}
\newcommand{\WA}{\texttt{WA}\xspace}
\newcommand{\ConfPass}{\texttt{confident\_pass}\xspace}
\newcommand{\ConfFail}{\texttt{confident\_fail}\xspace}
\newcommand{\Inconclusive}{\texttt{inconclusive}\xspace}
\newcommand{\Pass}{\texttt{pass}\xspace}
\newcommand{\Fail}{\texttt{fail}\xspace}

\newcommand{\Partial}{\textsc{PartCom}\xspace}
\newcommand{\Full}{\textsc{FullCom}\xspace}
\newboolean{showcomments}
\setboolean{showcomments}{true} 

\ifthenelse{\boolean{showcomments}}{
	\newcommand{\nbc}[3]{%
		{\colorbox{#3}{\bfseries\sffamily\scriptsize\textcolor{white}{#1}}}%
		{\textcolor{#3}{\sf\small$\langle$\textit{#2}$\rangle$}}%
	}
	
}{
	\newcommand{\nbc}[3]{}

}

\title{Noise-Resilient Testing of Quantum Programs Based on Output Dominance
}
\author[1]{Yuechen Li}
\author[1]{Yige Yang}
\author[1,*]{Tao Yue}
\affil[1]{Beihang University\\
\texttt{\{liyuechen, yangyige, yuetao\}@buaa.edu.cn}}
\affil[*]{Corresponding author}
\date{}

\begin{document}

\maketitle

\begin{abstract}
Quantum software testing (QST) is a central quality-assurance activity for checking whether quantum programs conform to their specifications. In the noisy intermediate-scale quantum (NISQ) era, gate, decoherence, and measurement errors distort observed outputs, making it difficult to distinguish program defects from noise. The existing noise-aware QST literature typically mitigates or models noise before applying a test oracle. However, these strategies can incur substantial cost and depend on calibration data, model training, or backend-specific information, limiting their applicability.
We present Qomino, a dominance-guided QST approach for judging tests from noisy samples without a separate noise-estimation or mitigation phase. The approach combines decomposition-aware inverse construction, comparison of the expected outcome with the strongest observed competitor, and bounded retries for inconclusive decisions.
We evaluate Qomino on 30 controlled buggy variants from six Qiskit programs and six simulated noisy backends, comparing it with six baselines spanning statistical, quantum-specific, and learning-based methods. Qomino achieves $85.65\%$ accuracy over all six programs and $99.95\%$ in the restricted two-program evaluation, significantly outperforming the corresponding baselines.
Across the evaluated programs and backends under the stated settings, Qomino is faster than the compared quantum-specific and learning-based methods. Ablation and retry analyses support the contributions of its main components.
\end{abstract}


\section{Introduction}\label{sec:intro}



Gate-based quantum computing uses qubits as basic units to process information. A quantum program specifies a sequence of quantum gates that produces the program's output state and the measurements that read out the result, where quantum measurement is the only means to observe outputs from a quantum system, whether executed on a real device or a quantum simulator.
As these quantum programs become larger and more integrated with classical workflows, quantum software engineering (QSE)~\cite{pezze20252030, murillo2025quantumsoftwareengineering, zhao2020quantum} addresses their development lifecycle, including phases such as testing, debugging, and maintenance.
Quantum software testing (QST)~\cite{paltenghi2024survey, li2026methodological, ali2023quantum, leite2025testing, miranskyy2019testing} is a central QSE activity that decides whether a quantum program satisfies its specification. Unlike its classical counterpart, a correct quantum program is not guaranteed to produce identical outputs across repeated executions of the same input, because measurement is probabilistic; a reliable approach must therefore separate this expected variation from bitstrings caused by a program defect, avoiding both false alarms on correct programs and missed bugs.

\looseness=-1
However, today's quantum computers remain in the noisy intermediate-scale quantum (NISQ) era~\cite{preskill2018nisq}: they execute programs before fault-tolerant error correction becomes available, and it is unclear when this phase ends. Near-term devices are subject to significant gate, decoherence, and measurement errors, which distort observed outputs far beyond the probabilistic variation of an ideal execution.
If QST ignores these effects, test approaches are hardly applicable in practice.
To address this problem, an emerging trend for noise-specific QST is either compensating for test outputs by estimating the noise~\cite{muqeet2024quantum} or mitigating noise before judging the test result~\cite{muqeet2024mitigatingno}.
However, both majorly recover the noise-free ideal output first, requiring extra executions, calibration, or pre-trained models, which raises testing cost and limits applicability when such characterization is unavailable. 
This motivates a different direction: rather than recovering the ideal output, we ask whether a test can resist noise effectively and return reliable pass/fail verdicts directly from noisy execution. 
Our key design \textit{hypothesis} is that, despite noise causing some shots to violate the specification, a correct program still produces its expected output as the \textit{dominant} outcome, which enables test-result decisions from noisy observations; without it, the intended output would be indistinguishable from noise.
 
To this end, we present \ApproachName, a novel QST approach that makes the expected output collapse to an ideally deterministic measurement result through an inverse operation, so that a single expected bitstring exists and its dominance can be assessed directly under noisy execution, without any separate noise-estimation or mitigation phase. \ApproachName realizes this through a combination of three components: 1) \textit{automatic state decomposition}, which partitions the specification state into separable parties so the inverse operation is built from smaller-scale operations, reducing circuit overhead and limiting noise amplification; 2) \textit{dominance analysis}, which compares the expected bitstring with the strongest competing outcome, while a Hoeffding-based decision quantifies whether the observed dominance is sufficient given finite-shot uncertainty; and 3) a \textit{bounded retry mechanism}, which collects additional samples when the decision is inconclusive, compensating statistically before a final verdict is returned. Together, these components let \ApproachName remain reliable under noise while keeping the execution cost affordable.

In fact, \ApproachName's dominance hypothesis is an operating assumption, not a guarantee for arbitrary noisy processes.
We therefore evaluate \ApproachName by reusing real-world scalable Qiskit programs and controlled test suites from a recent work~\cite{li2026dynamic}, plus six simulated noisy backends. We compare it against six baselines from three families (typical statistical methods, quantum-specific processing, and learning-based noise mitigation) on cost and effectiveness, together with ablation and retry analyses.
In the full-evaluation setting, \ApproachName achieves the highest overall accuracy of $85.65\%$, substantially exceeding all baselines, indicating that its verdicts largely agree with the ground truth. Its execution time is also lower than that of the state-of-the-art (SOTA) approaches designed for noisy QST, while ablations confirm that all three components contribute to the overall performance.

The main contributions of this paper include providing:
1) a dominance-guided, noise-resilient QST workflow that returns reliable test results directly from noisy executions, without recovering ideal outputs through a separate noise-estimation or mitigation phase;
2) an automatic state-decomposition procedure that builds smaller inverse operations from separable specification parties, reducing execution cost and limiting noise amplification from unnecessarily large circuits. 
3) a statistically controlled decision pipeline that combines Hoeffding-based confidence analysis with bounded retries to resolve inconclusive results within a limited shot budget. 
4) a systematic empirical study on real-world Qiskit programs and noisy backends, showing that \ApproachName achieves the highest effectiveness and executes faster than the compared SOTA approaches for noisy QST.

The remainder of this paper is organized as follows: Section~\ref{sec:background} introduces basics of quantum computing and noisy execution; Section~\ref{sec:methodology} presents \ApproachName; Section~\ref{sec:empirical} describes the experiment design and reports the results; Section~\ref{sec:threats} discusses threats to validity; and Section~\ref{sec:related} reviews related work, followed by the conclusion and future work in Section~\ref{sec:conclusion}.

\section{Background}\label{sec:background}
This section briefly reviews the quantum-computing concepts needed in this paper.

\textbf{Gate-based quantum programs.}
As a counterpart of bits in classical computation, qubits work as the units of quantum computation. A qubit is described by the normalized state vector $\ket{\psi}=\alpha\ket{0}+\beta\ket{1}$, where $\ket{0}$ and $\ket{1}$ are the computational-basis states, $\alpha,\beta\in\mathbb{C}$ are amplitudes, and $|\alpha|^2+|\beta|^2=1$. The joint state of multiple qubits is represented in the tensor-product space of their individual states, written $\ket{\psi_A}\otimes\ket{\psi_B}$ for two subsystems $A$ and $B$. A quantum circuit corresponding to a gate-based program generally has three stages: initial-state preparation transforms the all-zero state into the desired input; unitary evolution applies a sequence of reversible quantum gates, collectively represented by $U$ with $UU^\dagger=I$, to perform the computation; and measurement converts the resulting quantum state into classical data. For the one-qubit state $\alpha\ket{0}+\beta\ket{1}$ above, \textit{Pauli-Z measurement} returns $0$ and $1$ with probabilities $|\alpha|^2$ and $|\beta|^2$, respectively.
Pauli-Z measurement usually default in gate-based quantum programming refers to the observation in terms of computational basis, and the measurement results are associated with the basis selection.

\textbf{Quantum states.}
A quantum state is \emph{pure} when a single ket $\ket{\psi}$ completely describes it, and \emph{mixed} when it represents a statistical mixture of possible pure states. Mathematically, a pure state has $\rho=\ket{\psi}\bra{\psi}$ and $\operatorname{Tr}(\rho^2)=1$, whereas a mixed state is a weighted sum of pure-state density operators and satisfies $\operatorname{Tr}(\rho^2)<1$, where $\operatorname{Tr}(\cdot)$ is the trace operator in linear algebra, which adds the diagonal elements of a matrix and is used in quantum mechanics to describe properties such as state purity.
For a joint system $AB$ described by $\rho_{AB}$, the partial trace over $B$ removes $B$ and yields the reduced density operator $\rho_A=\operatorname{Tr}_{B}(\rho_{AB})$, preserving the statistics obtainable from $A$ alone. For a pure bipartite state, entanglement between subsystems $A$ and $B$ implies that the reduced state of either subsystem is mixed, whereas independent subsystems in a product pure state have pure reduced states. More generally, an $n$-qubit pure state is \textit{$k$-separable} when its qubits can be partitioned into $k$ mutually disjoint parties whose joint state is the tensor product of the individual quantum states of those subsystems; $k=n$ is \textit{fully separable}, while a state with no non-trivial separable partition is \textit{fully entangled}~\cite{dur2000classification, seevinck2008partial}.

\textbf{Quantum noise in the NISQ era.}
Current quantum computing is largely conducted in the NISQ era, in which available devices are not yet generally fault tolerant~\cite{preskill2018nisq}. Their behavior is inherently noisy because unavoidable physical effects, especially decoherence, perturb the intended circuit operations; imperfect control and readout add further errors. Compared with a noise-free (ideal) environment, these effects change the circuit's theoretical output distribution. Therefore, QST under noise is important for realistically validating executions on NISQ devices, and difficult because effects from program bugs and hardware noise can be mixed in the observed output, making their respective contributions hard to distinguish.

\section{Approach}\label{sec:methodology}
\begin{figure}
    \centering
    \includegraphics[width=0.98\textwidth]{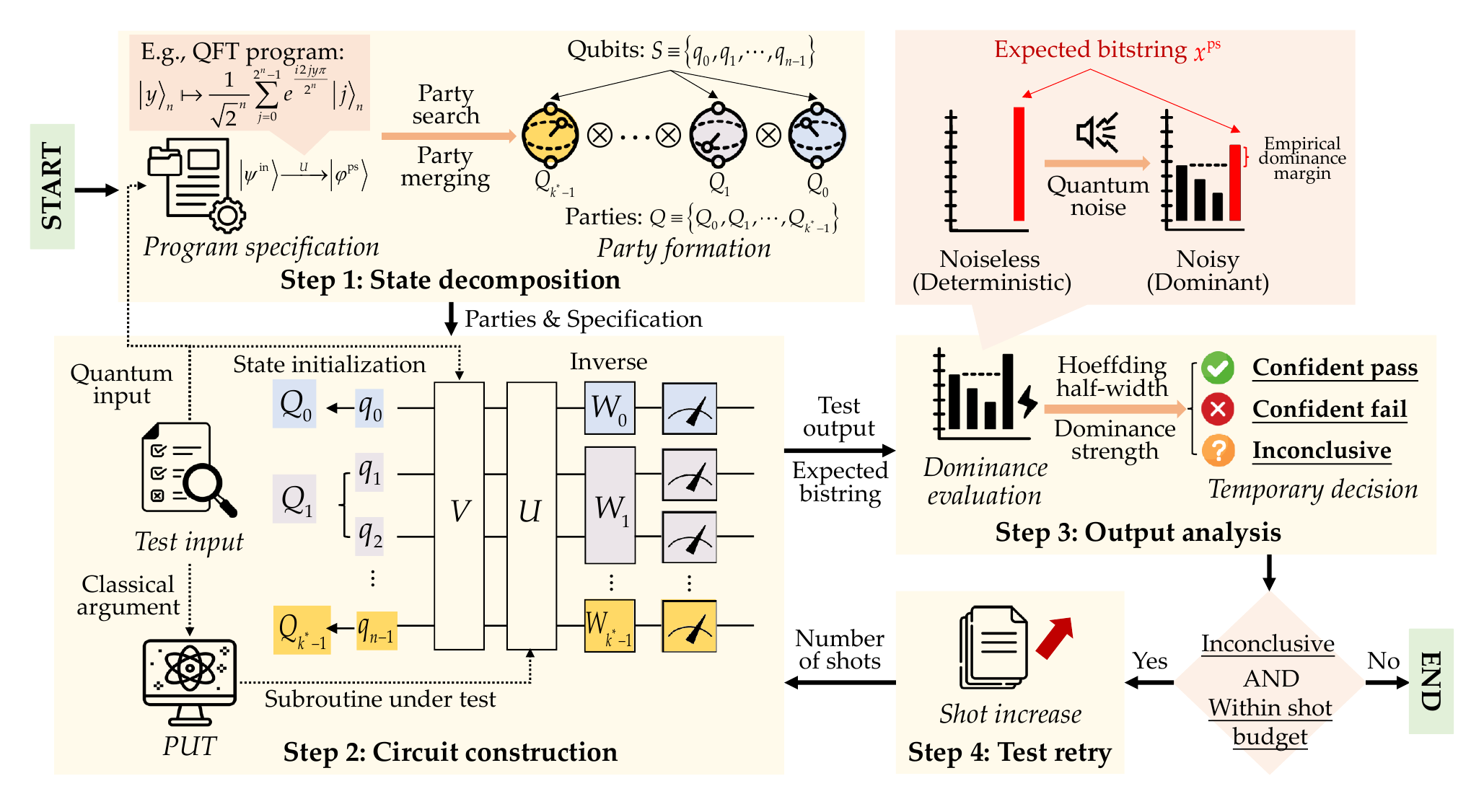}
    
    \caption{{Framework of \ApproachName}}
    
    \Description{A figure displaying the pipeline of the proposed approach Qomino.}
    \label{fig: pipeline}
\end{figure}

\ApproachName operates as a noise-oriented framework for black-box testing of quantum programs.
As illustrated in Figure~\ref{fig: pipeline}, it automatically  forms $k^*$ separable and disjoint parties, $Q \equiv \{Q_0, Q_1, \cdots, Q_{k^*-1}\}$, from the expected output state $\ket{\varphi^{\text{ps}}}$ of the
program specification that depicts how the initial state $\ket{\psi^{\text{in}}}$ maps to the specification $\ket{\varphi^{\text{ps}}}$.
\ApproachName then constructs an $n$-qubit quantum circuit by composing three quantum subroutines: state initialization $V$, the subroutine under test $U$, and an inverse operation $W$, where $U$ denotes the instantiated circuit of the program under test (PUT), and $W$ is defined as $W_{k^*-1}\otimes\cdots\otimes W_1\otimes W_0$.
Operation $W$ within the circuit enables the conversion of expected outputs from probabilistic into deterministic bitstrings in a noise-free environment, where $W_j$ is constructed according to the expected output of the corresponding party $Q_j$.
When testing on a noisy backend, \ApproachName instead analyzes whether the unique expected bitstring remains dominant over other outcomes that may arise from quantum noise or
program bugs.
Compared to the noiseless environment, a test in noise is considered successful only when the expected bitstring exhibits sufficient dominance.
If the evidence is inconclusive, \ApproachName may increase the number of shots within the configured budget and repeat the test; otherwise, it reports the corresponding pass or fail
result.

\subsection{State Decomposition}
\label{sec:decomposition}
In QST, state decomposition has been explored to reduce the circuit complexity when borrowing additional qubits and gates for testing~\cite{li2026dynamic}. Based on partial separability of the specification $\ket{\varphi^{\text{ps}}}$, this existing approach can form $k^*$ parties separable from each other. However, current implementation of decomposition is limited in generalization, because experts are required to manually read the program specification and accomplish the decomposition just based on an abstract formula. To deal with such limitations, we propose an automatic decomposition method from the concrete state $\ket{\varphi^{\text{ps}}}$ to such $k^*$ parties by search.

According to the principle of quantum information theory~\cite{nielsen2000quantum}, a subset of qubits can be separated from the remaining qubits when the two parts are independent, meaning that the overall state can be represented as the tensor product of their local states. Therefore, we recursively search for such bipartitions and use the purity of one part's reduced state as an operational test for this separability.
Let $S\equiv\{q_j\}_{j=0}^{n-1}$ denote the set of qubits covered by $\ket{\varphi^{\text{ps}}}$. Starting from this complete qubit set, \ApproachName recursively explores candidate bipartitions $\left(A,S\setminus A\right)$. For each candidate subset $A\subseteq S$, it computes the reduced density operator of the subsystem $A$,
\begin{equation}
    \rho_A=\operatorname{Tr}_{S\setminus A}\left(\ket{\varphi^{\text{ps}}}\bra{\varphi^{\text{ps}}}\right).
\end{equation}
Given a tolerance $\varepsilon_{\text{dec}} (\varepsilon_{\text{dec}}>0)$ due to the numerical error, the candidate partition is accepted when the reduced state is sufficiently pure, i.e.,
\begin{equation}\label{eq:purity_tolerance}
    \left|\operatorname{Tr}\left(\rho_A^2\right)-1\right|\leq \varepsilon_{\text{dec}}.
\end{equation}
This condition indicates that $A$ is separable from its complement and can therefore be searched independently. The search continues on the subset until no accepted bipartition remains. As a result, the specification is represented by $k$ mutually disjoint parties $\{Q_0,\ldots,Q_{k-1}\}$ that cover $S$ and satisfy  $\ket{\varphi^{\text{ps}}}=\ket{\varphi^{\text{ps}}_{k-1}}\otimes\cdots\otimes\ket{\varphi^{\text{ps}}_1}\otimes\ket{\varphi^{\text{ps}}_0}$, where $\ket{\varphi^{\text{ps}}_{j}}$ is denoted as the local expected output state of party $Q_j$. 

Inspired by Li et al.~\cite{li2026dynamic}, we further reduce the number of parties from $k$ into $k^*$ ($k\ge k^*$) by merging. 
This helps lower potential execution cost by grouping together the parties whose specification indicates deterministic measurement outcomes upon the selected measurement basis. 
\textbf{Assume that our scope is Pauli-Z measurement} and let $\{0,1\}^{|Q_j|}$ denote the measurement bitstrings over party $Q_j$, where in such measurement basis, the bitstring $x (x\in\{0,1\}^{|Q_j|})$ corresponds to the computational-basis state $\ket{x}$.  
We classify $Q_j$ as deterministic if there exists an output bitstring $x$ whose fidelity with the local specification state is at least $1-\varepsilon_{\mathrm{dec}}$, i.e.,
\begin{equation}
    \exists x\in\{0,1\}^{|Q_j|}:\quad F\!\left(\ket{x},\ket{\varphi^{\text{ps}}_j}\right)=\left|\langle x\mid\varphi^{\text{ps}}_j\rangle\right|^2\geq 1-\varepsilon_{\mathrm{dec}}.
\end{equation}
Otherwise, $Q_j$ is classified as non-deterministic. 
Note that, our clarification of deterministic or non-deterministic parties is relative to the selected measurement basis.
The tensor product of deterministic local states is concentrated on the bitstring formed by combining their expected outputs, so all deterministic parties can be merged into one party. Each non-deterministic party is retained separately. For example, a 4-qubit system with $3$ parties $\{Q_0, Q_1, Q_2\}$ and their specification states $\ket{\varphi^{\text{ps}}_0}_1=\ket{0}$, $\ket{\varphi^{\text{ps}}_1}_1=\ket{1}$, and $\ket{\varphi^{\text{ps}}_2}_2=\frac{1}{\sqrt{2}}\left(\ket{00}+\ket{11}\right)$\footnote{For rigor, the subscript $n$ of the notation $\ket{\cdot}_n$ highlights this state composed of $n$ qubits.} can be regrouped into $2$ parties $\{Q_{0'} , Q_{1'}\}$ with $\ket{\varphi^{\text{ps}}_{0'}}_2=\ket{\varphi^{\text{ps}}_1}_1\otimes\ket{\varphi^{\text{ps}}_0}_1=\ket{10}$ and $\ket{\varphi^{\text{ps}}_{1'}}_2=\ket{\varphi^{\text{ps}}_2}_2$, where the expected bitstring of the first party is deterministic while the second is probabilistic.

Algorithm~\ref{alg:state-decomposition} forms the final $k^*$ parties in two phases. In the first phase (Lines~\ref{alg:state-decomposition:phase1-start}--\ref{alg:state-decomposition:phase1-end}), it iteratively processes qubit subsets from the queue, accepts separable bipartitions according to the reduced-state purity criterion, and collects the subsets for which no further accepted split is found as the $k$ specification parties in $\mathcal{Q}_{\mathrm{spec}}$. Here, \textit{FindPureSplit} examines the nonempty proper subsets $A$ of the current subset $R$ in random order and returns the first subset for which the reduced states of $A$ and $R\setminus A$ satisfy the purity tolerance \eqref{eq:purity_tolerance}. It returns $\varnothing$ if no candidate satisfies the criterion. In the second phase (Lines~\ref{alg:state-decomposition:phase2-start}--\ref{alg:state-decomposition:phase2-end}), \textit{LocalState} returns the local pure state associated with each terminal subset $R\in\mathcal{Q}_{\mathrm{spec}}$. After classification, \textit{MergeDeterministic} returns a singleton containing the merged deterministic party, or the empty set if none exists. The algorithm forms the final execution-party set by taking the union of this result and the retained non-deterministic parties.

\begin{algorithm}[t]
    \caption{Automatic State Decomposition via Bipartition Search}
    \label{alg:state-decomposition}
    {
    \begin{algorithmic}[1]
        \Require Specification state $\ket{\varphi^{\text{ps}}}$, qubit set $S\equiv\{q_j\}_{j=0}^{n-1}$, error limit $\varepsilon_{\mathrm{dec}}$
        \Ensure A set of $k^*$ mutually disjoint execution parties $Q=\{Q_0,\ldots,Q_{k^*-1}\}$
        \State $\mathcal{R}\leftarrow[S]$ \Comment{Initialize the subset queue} \label{alg:state-decomposition:phase1-start}
        \State $\mathcal{Q}_{\mathrm{spec}}\leftarrow\varnothing$ \Comment{Collect specification parties}
        \While{$\mathcal{R}$ is not empty}
            \State $R\leftarrow\textbf{Dequeue}(\mathcal{R})$ \Comment{Remove the next subset}
            \State $A\leftarrow\textbf{FindPureSplit}(\ket{\varphi^{\mathrm{ps}}},R,\varepsilon_{\mathrm{dec}})$ \Comment{Return the first subset whose two reduced states satisfy the purity criterion}
            \If{$A\neq\varnothing$}
                \State $\mathcal{R}\leftarrow\textbf{Enqueue}(\mathcal{R},A)$ \Comment{Continue decomposing $A$}
                \State $\mathcal{R}\leftarrow\textbf{Enqueue}(\mathcal{R},R\setminus A)$ \Comment{Continue decomposing $R\setminus A$}
            \Else
                \State $\mathcal{Q}_{\mathrm{spec}}\leftarrow\mathcal{Q}_{\mathrm{spec}}\cup\{R\}$ \Comment{No candidate split satisfies the purity criterion}
            \EndIf
        \EndWhile \label{alg:state-decomposition:phase1-end}
        \State $\mathcal{D}\leftarrow\varnothing$ \Comment{Collect deterministic parties} \label{alg:state-decomposition:phase2-start}
        \State $\mathcal{N}\leftarrow\varnothing$ \Comment{Collect non-deterministic parties}
        \ForAll{$R\in\mathcal{Q}_{\mathrm{spec}}$}
            \State $\ket{\varphi_R^{\mathrm{ps}}}\leftarrow\textbf{LocalState}(\ket{\varphi^{\mathrm{ps}}},R)$ \Comment{Extract the local state}
            \State $x_R\leftarrow\mathop{\arg\max}_{x\in\{0,1\}^{|R|}}\left|\langle x\mid\varphi_R^{\text{ps}}\rangle\right|^2$ \Comment{Select the most probable bitstring}
            \If{$\left|\left|\langle x_R\mid\varphi_R^{\text{ps}}\rangle\right|^2-1\right|\leq\varepsilon_{\mathrm{dec}}$}
                \State $\mathcal{D}\leftarrow\mathcal{D}\cup\{(R,x_R)\}$ \Comment{Record a deterministic party}
            \Else
                \State $\mathcal{N}\leftarrow\mathcal{N}\cup\{(R,\ket{\varphi_R^{\text{ps}}})\}$
            \EndIf
        \EndFor
        \State $Q\leftarrow\textbf{MergeDeterministic}(\mathcal{D})\cup\mathcal{N}$ \Comment{Form the execution parties}
        \State Return $Q$ \label{alg:state-decomposition:phase2-end}
    \end{algorithmic}
    }
    
\end{algorithm}

\subsection{Circuit Construction}\label{sec:circuit_construction}
After obtaining $k^*$ parties, \ApproachName first constructs one test circuit by appending the state initialization subroutine for $\ket{\psi^{\text{in}}}$ (i.e., $\ket{\psi^{\text{in}}}=V\ket{0}^{\otimes n}$) and the subroutine under test $U$ determined by classical arguments of the test input. 
{Note that gate-based quantum computation is generally initialized with a simple state $\ket{\psi^{\text{in}}}$, whereas a more complex $\ket{\psi^{\text{in}}}$ with high entanglement is often considered in measurement-based quantum computation~\cite{raussendorf2001one,raussendorf2003measurement}.
This enables us to be more focused on easing preparation of the specification state $\ket{\varphi^{\text{ps}}}$ instead of the initial state $\ket{\psi^{\text{in}}}$ when building a quantum circuit.
For example, Quantum Fourier Transform (QFT) shown in Figure~\ref{fig: pipeline} starts with computational-basis states $\ket{y}_n$ ($y\in\{0,1\}^{n}$) and ends with $\frac{1}{\sqrt{2}^n}\sum_{j=0}^{2^n-1}e^{2\pi i y j/2^n}\ket{j}_n$ involving both superposition and related phases. We can implement $V$ by applying Pauli-$X$ gates to selected qubits to encode $y$ (e.g.,
$X\otimes I\otimes X$ for $101$) or equivalently by leveraging Qiskit's \texttt{StatePreparation} API for preparation based on its statevector (e.g., $[0,0,0,0,0,1,0,0]^{\top}$) for $101$.}

After that, \ApproachName inserts a party-specific inverse operation $W$ at the end of $U$, and then measures the involved qubits. The inverse operation is inspired by existing studies~\cite{long2024testingmultisubroutine,long2025blackbox} to enable checking the ideal circuit outcomes as deterministic bitstrings rather than analyzing complex probabilistic distributions.
Nevertheless, they usually introduce a high-dimensional inverse operation without the decomposition into basic quantum gates that operate on fewer qubits. This may raise simulation cost and amplify quantum noise by significantly increasing the circuit size and depth.

Instead, the inverse operation in \ApproachName depends on the type of each party, with $W$ decomposed into $k^*$ components $\left\{W_j\right\}_{j=0}^{k^*-1}$. For a deterministic party $Q_j$, the local specification $\ket{\varphi_{j}^{\text{ps}}}$ is a computational-basis state 
for our discussed Pauli-Z measurement, so no inverse operation is required (or denoted as $W_j \equiv I$) and \ApproachName directly records the expected bitstring $x^{\text{ps}}_j$ as $\left(\varphi_{j}^{\text{ps}}\right)_{\mathrm{b}}$. Compared to these existing studies, \ApproachName leverages the state property to avoid unnecessary quantum gates in inverse. For a non-deterministic party $Q_j$, \ApproachName appends an inverse of its specification-state preparation to achieve $x^{\text{ps}}_j$ as $0\cdots 0$ with the length of $\left|Q_j\right|$.
Specifically, for each non-deterministic party $Q_j$, \ApproachName first obtains a unitary operator $P_j$ for preparing $\ket{\varphi_j^{\text{ps}}}$, i.e., 
$P_j\ket{0}^{\otimes |Q_j|}=\ket{\varphi_j^{\text{ps}}}$,
where $P_j$ is also practically available via Qiskit's \texttt{StatePreparation} API from the specific statevector of $\ket{\varphi_j^{\text{ps}}}$.
Then, corresponding inverse operation is the Hermitian conjugate $W_j\equiv P_j^{\dagger}$, which maps the expected local state to the all-zero state:
\begin{equation}
    W_j\ket{\varphi_j^{\text{ps}}}=P_j^{\dagger}\ket{\varphi_j^{\text{ps}}}=\ket{0}^{\otimes |Q_j|}.
\end{equation}
For example, given parties $\{Q_{0'},Q_{1'}\}$ with $\ket{\varphi^{\text{ps}}_{0'}}_2=\ket{10}$ and $\ket{\varphi^{\text{ps}}_{1'}}_2=\frac{1}{\sqrt{2}}\left(\ket{00}+\ket{11}\right)$, the local inverse is only placed on $Q_1'$. The expected output bitstring $x^{\text{ps}}$ for this test is therefore $0010$, assuming qubits with higher indices appear on the left in the binary representation.  

\subsection{Output Analysis}
Applying $W$ provides an unambiguous assertion for deriving the test result in a noise-free environment: all repeated measurements must return $x^{\mathrm{ps}}$, and the occurrence of any other bitstring indicates a failure. This exact-match criterion cannot be directly transferred to a noisy backend, where even a correct quantum program may produce occasional non-$x^{\mathrm{ps}}$ outcomes because of quantum noise. 
Instead of a provable theorem, we make a \textbf{conservative hypothesis} that \textit{a practically usable correct PUT should nevertheless preserve $x^{\mathrm{ps}}$ as an outcome that is sufficiently distinguishable from noise-induced alternative bitstrings on NISQ hardware}. If noise overwhelms the intended output, the hardware execution may provide little use for practical quantum applications. We therefore introduce output dominance as the noise-aware criterion, evaluating whether $x^{\mathrm{ps}}$ remains sufficiently dominant over the strongest observed competitor rather than requiring every measurement to match it.

Let $s$ be the number of shots used in one check and $c_{\text{ps}}$ be the number of samples producing $x^{\mathrm{ps}}$. We call each observed bitstring other than $x^{\mathrm{ps}}$ a \textit{competitor}, and denote $c_{\text{com}}$ as the number of samples producing \textbf{the most frequently observed competitor} $x^{\mathrm{com}}$. 
Note that $x^{\text{ps}}$ and $x^{\text{com}}$ should be discussed in the same cumulative count map of the $s$ shots.
We then define the empirical dominance margin as
\begin{equation}
    d\equiv\frac{c_{\text{ps}}-c_{\text{com}}}{s}.
\end{equation}
The margin compares the expected output with the most frequent competitor rather than with the aggregate count of all incorrect outcomes, because a single alternative outcome occurring as frequently as the expected output can already challenge its dominance. It therefore captures the strongest directly observable competing explanation for the execution result.

The observed counts provide empirical frequencies rather than the underlying outcome probabilities. Let $p_{\text{ps}}$ and $p_{\text{com}}$ denote the underlying probabilities of $x^{\mathrm{ps}}$ and $x^{\mathrm{com}}$, respectively; their empirical estimates are $c_{\text{ps}}/s$ and $c_{\text{com}}/s$, whose difference is the observed margin $d$. To account for finite-shot uncertainty in this empirical margin, we use a Hoeffding-based half-width~\cite{hoeffding1963}. For a fixed specification-versus-competitor comparison and independent, identically distributed shots, define $\mathcal{Z}_j=1$ for the expected outcome, $\mathcal{Z}_j=-1$ for that competitor, and $\mathcal{Z}_j=0$ otherwise. Let $\overline{\mathcal{Z}}=\frac{1}{s}\sum_{j=1}^{s}\mathcal{Z}_j$, and then
\begin{equation}
     \mathbb{E}\left[\overline{\mathcal{Z}}\right]
    =\mathbb{E}\!\left[\frac{1}{s}\sum_{j=1}^{s}\mathcal{Z}_j\right]
    =\frac{1}{s}\sum_{j=1}^{s}\mathbb{E}[\mathcal{Z}_j]
    =p_{\text{ps}}-p_{\text{com}},
\end{equation}
where the last equality uses the identical distribution of the shots, whereas the empirical mean is $\overline{\mathcal{Z}}=d$. For these independent random variables in $[-1,1]$, Hoeffding's inequality states that
\begin{equation}
    \label{eq:hoeffding}
    \Pr\left(\left|\overline{\mathcal{Z}}-\mathbb{E}\left[\overline{\mathcal{Z}}\right]\right|\geq t\right)
    \leq 2 e^{ -\frac{st^2}{2} }.
\end{equation}
Based on \eqref{eq:hoeffding}, we use the half-width obtained by setting the right-hand side to the per-evaluation error budget $\delta$ for the observed margin:
\begin{equation}
    \label{eq:half_width}
    t=\sqrt{\frac{2}{s}\ln \frac{2}{\delta}}.
\end{equation}
Thus, $d$ is the center of the uncertainty range used for the dominance decision, while $d-t$ and $d+t$ are its lower and upper bounds, respectively. This interval is used as a confidence-guided rule for the selected comparison rather than establishing a guarantee over all possible competitors.

Dominance is expected as a property of the underlying outcome probabilities, whereas finite-shot execution provides only the empirical margin $d$. Thus, we adopt $\gamma$ as the minimum dominance strength configured for the test and compare the uncertainty range around $d$ with this threshold. 
Especially, $\gamma$ should be non-negative to support a minimum dominance of the expected $x^{\mathrm{ps}}$ over a fixed $x^{\mathrm{com}}$, and its maximum must be constrained by the probability sum as one.
Then, $\tilde{r}$ is introduced as a temporary decision of the test result, which may remain provisional rather than immediately yielding a final pass/fail:
\begin{equation}
    \tilde{r}\equiv
    \begin{cases}
        \ConfPass, & d-t>\gamma,\\
        \ConfFail, & d+t<\gamma,\\
        \Inconclusive, & d-t\leq\gamma\leq d+t.
    \end{cases}
\end{equation}
The three states distinguish whether the uncertainty range lies above the dominance threshold, below it, or overlaps it. In particular, $d-t>\gamma$ indicates that the expected output remains dominant with certain sufficiency even after accounting for finite-shot fluctuation, while $d+t<\gamma$ indicates that the dominance strength is not reached even after the empirical margin is increased by $t$ to account for finite-shot uncertainty. If $d-t\leq\gamma\leq d+t$, the evidence is inconclusive in view of statistical evidence. 
As for the final binary test result $r\in\{\Pass,\Fail\}$, the temporary decision $\tilde{r}$ can be assigned to $r$ when the evidence is confident: \ConfPass and \ConfFail correspondingly yield $r=\Pass$ and $r=\Fail$. If $\tilde{r}$ remains \Inconclusive, no final binary result is assigned at this stage, and the retry procedure described in the next subsection will be invoked.

\subsection{Test Retry}
In practice, returning an inconclusive outcome provides limited value to testers because it leaves the testing process incomplete. Inspired by a recent QSE study~\cite{luo2026qemi}, we introduce a retry stage in \ApproachName that incrementally increases the number of shots within a configured budget. Beyond simply determining on the closeness between test and referred outputs in that work, our retry mechanism resolves statistical inconclusiveness based on the Hoeffding bound and a retry-aware allocation of the error probability across evaluations. More specifically, when the initial execution yields $\tilde{r}=\Inconclusive$, \ApproachName collects additional shots, merges their counts with the cumulative counts, and recomputes the decision from all samples. The growing sample size reduces the Hoeffding half-width $t$, while $\Inconclusive$ remains only an intermediate trace and is never returned as the final result.

Determined by the test configuration, the initial execution uses $s_0$ shots that can be configured by users, and the additional shots are limited by a retry budget $b_{\text{retry}}$. For the $j$-th retry, where $j\geq1$, the planned batch size is $s_0 2^{j-1}$. When the collected samples are far from sufficient for a confident decision, this geometric schedule avoids many small retries, thereby reducing the number of repeated circuit executions. Denoting the remaining budget before the $j$-th retry by $B_{j-1}$, the actual batch size and updated budget are
\begin{equation}
    \widetilde{s}_j=\min\left\{s_0 2^{j-1},B_{j-1}\right\},
    \label{eq:retry-schedule}
\end{equation}
where $B_j=B_{j-1}-\widetilde{s}_j$ and $B_0=b_{\text{retry}}$. The maximum number of retries is
$N_{\mathrm{retry}}=\lceil\log_2(1+b_{\mathrm{retry}}/s_0)\rceil$,
resulting in at most $N_{\mathrm{retry}}+1$ evaluations including the initial execution.

\begin{algorithm}[t]
    \caption{Finite-Budget Test Retry in \ApproachName}
    \label{alg:test-retry}
    {
    \begin{algorithmic}[1]
        \Require Test circuit \texttt{qc}, initial counts $\mathcal{C}_0$, expected output $x^{\mathrm{ps}}$, dominance threshold $\gamma$, initial shots $s_0$, retry budget $b_{\text{retry}}$, user-configured error probability $\delta_{\mathrm{user}}$, initial random seed $seed\in\mathbb{Z}\cup\{\texttt{None}\}$
        \Ensure A binary test result $r\in\{\Pass,\Fail\}$  
        \State $N_{\mathrm{retry}}\leftarrow\lceil\log_2(1+b_{\text{retry}}/s_0)\rceil$
        \State $\mathcal{C}\leftarrow\mathcal{C}_0$ \Comment{Record measurement counts from the initial execution}
        \State $B_0\leftarrow b_{\text{retry}}$
        \For{$j \in  \{1,2,\cdots,N_{\mathrm{retry}}\}$}
            \State $\widetilde{s}_j\leftarrow\min\{s_0 2^{j-1},B_{j-1}\}$
            \State $seed_j\leftarrow
            \begin{cases}
                \texttt{None}, & seed=\texttt{None},\\
                seed+j, & \text{otherwise}
            \end{cases}$ \Comment{Follow a traceable random stream if $seed$ is specified}
            \State $\mathcal{C}_j\leftarrow\textbf{TestCircuit}(\texttt{qc},\widetilde{s}_j,seed_j)$ \Comment{Obtain measurement counts from the $j$-th retry}
            \State $\mathcal{C}\leftarrow\textbf{MergeCounts}(\mathcal{C},\mathcal{C}_j)$ \Comment{Accumulate counts}
            \State $s_j\leftarrow\sum_x\mathcal{C}(x)$ \Comment{Count cumulative shots}
            \State $B_j\leftarrow B_{j-1}-\widetilde{s}_j$ \Comment{Update the remaining retry budget}
            \State $\delta_j\leftarrow\delta_{\mathrm{user}}/(N_{\mathrm{retry}}+1)$
            \State $\tilde{r}\leftarrow\textbf{AnalyzeOutput}(\mathcal{C},s_j,x^{\mathrm{ps}},\gamma,\delta_j)$
            \If{$\tilde{r}=\ConfPass$}
                \State Return $r\leftarrow\Pass$  
            \ElsIf{$\tilde{r}=\ConfFail$}
                \State Return $r\leftarrow\Fail$  
            \EndIf
        \EndFor
        \State Return $r\leftarrow\Fail$ 
    \end{algorithmic}
    }
\end{algorithm}

Algorithm~\ref{alg:test-retry} summarizes the test process of \ApproachName invoking the retry stage, where the initial execution uses $s_0$ shots and returns a measurement-count map $\mathcal{C}_0$.
The initial execution is evaluation $j=0$ before triggering the retry, while evaluation $j\geq1$ follows the $j$-th retry whose implementation depends on the historical result decision. For each $j\in\{0,\ldots,N_{\mathrm{retry}}\}$, the corrected error probability is
$\delta_j=\delta_{\mathrm{user}}/(N_{\mathrm{retry}}+1)$.
Here, the user-configured hyperparameter $\delta_{\mathrm{user}}$ specifies the total error budget, whereas $\delta_j$ is the per-evaluation budget passed to the output analysis. 
Dividing this budget equally across the scheduled $N_{\mathrm{retry}}+1$ evaluations gives $\sum_{j=0}^{N_{\mathrm{retry}}}\delta_j=\delta_{\mathrm{user}}$ and should be interpreted as a conservative allocation of the user-specified error budget, rather than as a formal simultaneous guarantee over all adaptive paths.
For a reproducible simulation, each execution uses a fresh random stream to obtain independent samples. The procedure stops once the cumulative interval supports a confident pass or fail. 
If the retry budget is exhausted while the result remains inconclusive, \ApproachName returns a \Fail because the available evidence cannot establish sufficient dominance of the expected output, which violates the necessity of a passed test discussed in \ApproachName; such a consideration is driven by intuition and will be investigated in Section~\ref{sec:rq4}. 
Unlike a \ConfFail, the above \Fail does not indicate that the dominance interval lies below $\gamma$.

\section{Empirical Study}\label{sec:empirical}

We first formulate four research questions (RQs) (Section~\ref{sec:rq}), then describe the benchmark (Section~\ref{sec:benchmark}), the baselines and ablation variants used for comparison (Section~\ref{sec:baselines}), the evaluation metrics and statistical tests (Section~\ref{sec:metrics}), and the execution settings (Section~\ref{sec:settings}). Section~\ref{sec:experimental_results} then presents the experimental results answering each RQ.

\subsection{Research Questions}\label{sec:rq}
We design the four RQs to evaluate \ApproachName from complementary perspectives, as listed below:
\begin{enumerate}[label=\textbf{RQ\arabic*}, leftmargin=*]
	\item (\textsc{Effectiveness comparison}) How effective is \ApproachName in producing reliable test results compared to existing QST approaches?

    \item (\textsc{Cost comparison}) What is the time cost of executing \ApproachName for QST compared to existing QST approaches?

    \item (\textsc{Ablation study}) How does each component of \ApproachName contribute to the overall performance of QST?

    \item (\textsc{Retry investigation}) How does the retry mechanism behave under noisy execution, and how does its final fallback decision affect test-result reliability?
\end{enumerate}
RQ1 examines whether \ApproachName's test decisions agree with the ground truth, which is the primary criterion for judging effectiveness, as a reliable approach should correctly distinguish passing and failing tests rather than simply report more failures.
RQ2 investigates the execution cost, which is important because noise simulation itself is more costly than ideal simulation, and the QST approaches are not expected to additionally introduce excessive overhead. 
RQ3 isolates the contributions of the main components through controlled ablations, clarifying which design choices support effectiveness and cost. Finally,
RQ4 examines how often noisy executions require retries and whether the final fallback decision after the retry budget is exhausted affects test results. 

\subsection{Benchmark}\label{sec:benchmark}
PUTs and test suites are taken from the benchmark of the latest work~\cite{li2026dynamic},which focuses on the reliability of test results similar to ours.
In detail, the benchmark includes six real-world scalable Qiskit quantum programs, i.e., Linear Pauli Rotations (\LPR), Linear Amplitude Function (\LAF), Quantum Fourier Transform (\QFT), Diagonal Operation (\DO), Integer Comparator (\IC), and Weighted Adder (\WA), and 30 buggy mutants generated in a controlled manner.
Each PUT is paired with a test suite that contains 50 test cases labelled with ground truth and deliberately sets the proportion of failed test as 50\% to mitigate the bias in a controlled experimental evaluation.
Per program, the benchmark offers a program specification for classically calculating expected output states based on the real-world documentation, partially supporting that the required specification states are available in practice, and also enabling a functionality-grounded comparison between the actual and expected test outputs.

Aside from quantum inputs directly determining the operation $V$ in circuits, each test case fixes the classical arguments that instantiate a single quantum circuit (Section~\ref{sec:circuit_construction}), so our empirical study covers 30 mutants $\times$ 50 test cases $= 1{,}500$ circuits in total.
When being fully decomposed into basic gates and transpiled under ideal simulation, all contain 10 qubits and the maximum circuit gates and depths reach 4,179 and 4,117, which supports scalability evaluation of \ApproachName.

\subsection{Baselines and Variants}\label{sec:baselines}
Because an exhaustive comparison with all related QST studies is infeasible, we consider six state-of-the-art or representative approaches from three families, covering complementary and heterogeneous testing principles under a common benchmark:
\begin{itemize}[leftmargin=*]
    \item \textbf{Typical statistical methods}: These have been used in the majority of existing QST studies~\cite{li2026methodological}, as they are simple for comparing probability distributions after measurement with their references. We select the Chi-square test (\ChiSquare) and Jensen-Shannon divergence (\JSDiv) representative of hypothesis testing and statistical distance, respectively, owing to their popularity.
    \item \textbf{Quantum-specific processing methods}: We include three state-of-the-art approaches: \QOPS~\cite{muqeet2024quantum}, \DOSS~\cite{li2026dynamic}, and inverse test (\Inverse)~\cite{miranskyy2025cost,miranskyy2025feasibility}, where they introduce extra processing to match the characteristics of quantum programs. Other than directly measuring the original final states for analysis, they exploit extra quantum gates, qubits, or measurement transform to ensure test quality. Especially, \QOPS and \DOSS have already been evaluated in the noisy environment.
    \item \textbf{Learning-based noise mitigation methods}: The noise-oriented approach \QOIN~\cite{muqeet2024mitigatingno} is considered, since it pioneers leveraging machine learning to filter quantum noise involved in QST. Our evaluation for \QOIN simply loads the provided ready-trained models associated with specific programs and backends, and does not replicate its model training procedure. Thus, \QOIN is evaluated merely on \QFT and \WA, the overlap in programs' general functionalities between our selected benchmark and the programs used in \QOIN's original paper.
\end{itemize}
Because running all six baselines at full scale is prohibitively expensive, we split them into two groups considering the execution costs and application constraints: (1) \Partial: \QOIN and \QOPS, whose per-test execution time exceeds our budget (i.e., 20 seconds) or the approach itself can be merely adapted to certain programs, where this budget is reasonable for testing just 10-qubit circuits, as even one full execution of each approach under this budget would reach $20$ seconds $\times$ $1,500$ test cases $\times$ $6$ backends $= 50$ hours.
Hence, we evaluate them only on \QFT and \WA as strictly  constrained by \QOIN; (2) \Full: the remaining four approaches, which fit within the budget and are evaluated on all programs.

Also, we introduce four ablation variants independently implemented for addressing RQ3:
\begin{itemize}[leftmargin=*]
    \item \textbf{For state decomposition}: Variant \NoDecomposition disables the decomposition by using a full inverse operation $W$, while \NoBipartition further studies whether the decomposition relies on a specific search algorithm by replacing the bipartition (Section~\ref{sec:decomposition}) with random search.
    \item \textbf{For dominance evaluation}: Both \NoDominance and \NoHoeffding investigate how the output analysis of \ApproachName impacts the reliability of test results, where the former requires every measurement of a passed test to match the expected output (i.e., excluding the dominance), and the latter directly compares the empirical margin over all incorrect outcomes combined with $\gamma$ (i.e., $d>\gamma$ as \Pass; otherwise, as \Fail), without the Hoeffding identifier into three-candidate temporary decisions $\tilde{r}$ along with the test retry to deal with $\tilde{r}=$\Inconclusive.
\end{itemize}
RQ4 examines retry frequency and the reliability of the final decision when retries do not resolve the uncertainty. We therefore compare \ApproachName, which conservatively reports $r=\Fail$ in this situation, with the \InconclusiveToPass variant, which reports $r=\Pass$; the variant changes only this fallback decision, allowing us to isolate its effect.
 
\subsection{Evaluation Metrics}\label{sec:metrics}
\textbf{Cost and effectiveness metrics.} 
For cost, we define the total execution time of an approach for a single seed as the cumulative wall-clock time required to execute all test cases of all PUTs on all backends included in the discussed evaluation scope.
For effectiveness, instead of recording the number of reported faults, we measure the consistency between the actual test results and those provided by the ground truth, because the so-called faults may not be triggered by program bugs but misinterpreted attributed to quantum measurement and noise. To this end, we follow recent studies~\cite{li2026dynamic,muqeet2024quantum}, and adapt widely used metrics from binary classification, including accuracy, precision, recall, and F1-score~\cite{sokolova2009systematic}, where a positive and a negative correspond to the failure and success of a test, respectively.
We report both the mean and sample standard deviation of each metric over the repeated-seed runs within each reported scope, where the deviation is computed based on samples produced by different random seeds.

\textbf{Statistical tests.} 
All the statistical tests are based on samples independently generated by repeated seeds.
We compare each baseline approach or ablation variant with \ApproachName using the two-sided Mann--Whitney $U$ test on the seed-level samples. We report the resulting $p$-value together with the Vargha--Delaney effect size $\hat{A}_{12}$~\cite{vargha2000critique}, which quantifies the probability that a randomly selected observation from one group exceeds one from the other group. We use $p\text{-value}<0.05$ as the threshold for statistical significance and interpret the magnitude of $\hat{A}_{12}$ using the conventional negligible, small, medium, and large categories. 

\subsection{Execution Settings}\label{sec:settings}
\textbf{Approach parameters.} 
For a fair and controlled comparison, the decision thresholds (e.g., $p$-value for \ChiSquare) of both statistical methods are set as 0.05. Approaches from existing publications (i.e., \QOIN, \QOPS, and \DOSS) are configured with their recommended parameters in the original papers or the default in the open-source artifacts.
Regarding \ApproachName: the error bound is $\varepsilon_{\text{dec}}=10^{-10}$ to keep reasonably small; we select widely-used error probability $\delta_{\text{user}}=0.05$; the dominance strength is $\gamma=0$ to ensure its minimal role and avoid parameter tuning in our experiments; and the retry budget is $b_{\text{retry}}=1$ motivated by affordable cost in execution and fair comparison to approximate other approaches with only $s_0$ shots. 
These hyperparameters are held fixed to avoid method-specific tuning inside the comparison.

\textbf{Shots and repetition.}
For each approach, we choose $s_0=1,024$ default in Qiskit as the pre-configured number of shots for each circuit run. For the consideration of both statistical significance and execution cost, we repeat each circuit run for one test case with 10 different random seeds for approaches and variants except for \QOPS, while \QOPS is merely executed \textbf{once} due to its extremely high time cost (more details in Section~\ref{sec:cost_comparison}), thereby no variance claim on \QOPS in our experiments. Note that the configured seeds just refer to the initial $seed$ for \ApproachName (Algorithm~\ref{alg:test-retry}).

\textbf{Execution environment.}
We use Qolumbina~\cite{li2026benchmarking} as a dependency in our codebase, which provides QST-specific infrastructure built on Qiskit 2.3.0. 
We consider six \textit{fake} backends (\texttt{fake\_almaden}, \texttt{fake\_boeblingen}, \texttt{fake\_johannesburg}, \texttt{fake\_singapore}, \texttt{fake\_cairo}, and \texttt{fake\_cambridge}) to simulate quantum noise and hardware characteristics.
These backends are selected because they are supported by both Qolumbina and \QOIN, and their allowed maximum qubits range in $20$--$28$, just greater than those required by the PUTs while still keeping simulation time affordable.
All six are discussed across RQ1, RQ2 and RQ4, while the ablation study in RQ3 merely uses \texttt{fake\_almaden} to reduce the overall execution cost, where this backend has the smallest maximally allowed qubits among the six, and RQ1's empirical result indicates this backend along with three of the others exhibit similar impact on \ApproachName effectiveness (Section~\ref{sec:effectiveness_comparison}).
All experiments have been conducted on Ubuntu 22.04.5 LTS with AMD Ryzen Threadripper PRO 7975WX 32-Cores.

\subsection{Experimental Results}\label{sec:experimental_results}
In the following section, we answer each RQ.

\subsubsection{RQ1: Effectiveness Comparison}\label{sec:effectiveness_comparison}

Table~\ref{tab: rq1_overall} summarizes the effectiveness of all approaches over the evaluated backends and PUTs. Recall from Section~\ref{sec:baselines} that the \Full group covers all six programs, while the \Partial group (constrained by \QOIN's training data) covers only \QFT and \WA; \ApproachName is evaluated in both groups for comparison. In the \Full group, \ApproachName achieves the highest accuracy ($85.65\%$), precision ($86.75\%$), and F1-score ($85.60\%$), while its recall is $84.84\%$.
These results indicate that \ApproachName provides the most balanced agreement with the ground truth: it avoids the strong tendency of the other approaches to label nearly every test as a failure, which yields recalls close to $100.00\%$ but accuracies between $50.01\%$ and $65.03\%$.
For \Partial group with a narrower scope, \ApproachName achieves near-perfect results across all four metrics.
\QOIN and \QOPS reach $61.69\%$ and $49.97\%$ accuracy, with high recalls ($99.71\%$ and $94.07\%$), showing the same fail-leaning tendency.
Despite high recall reproduced, this differs from higher general effectiveness reported in their original studies~\cite{muqeet2024quantum, muqeet2024mitigatingno} (e.g., up to $1.000$ precision, recall, and F1-score for \QOPS), likely because our reused benchmark fixes the failure proportion at $50\%$ per program (Section~\ref{sec:benchmark}), so the metrics reflect agreement with ground truth without either class dominating the scores.

When looking at the results of the statistical tests, we can observe that: in the \Full group, all four baselines differ significantly from \ApproachName on every metric ($p\text{-value}<0.001$). For accuracy, precision, and F1-score, $\hat{A}_{12}=0.00$ for every baseline, indicating that their values are consistently lower than \ApproachName's; for recall, $\hat{A}_{12}=1.00$, reflecting the baselines' consistently higher recall and fail-leaning tendency. In the \Partial group, \ApproachName achieves significantly higher values than \QOIN on all four metrics ($p\text{-value}<0.05$, $\hat{A}_{12}\leq0.17$), although the absolute recall improvement is small (0.19 percentage points over \QOIN). 

\begin{table}[htbp]

    \caption{Approach effectiveness across all backends and programs}\label{tab: rq1_overall}
    
    \begin{center}
        \resizebox{1\columnwidth}{!}{
            \begin{tabular}{c|c|ccc|ccc|ccc|ccc}
\toprule[1pt]
\textbf{Group} & \textbf{Approach} & \multicolumn{3}{c|}{\textbf{Accuracy (\%)}} & \multicolumn{3}{c|}{\textbf{Precision (\%)}} & \multicolumn{3}{c|}{\textbf{Recall (\%)}} & \multicolumn{3}{c}{\textbf{F1-score (\%)}} \\
\cmidrule(lr){3-5}\cmidrule(lr){6-8}\cmidrule(lr){9-11}\cmidrule(lr){12-14}
 &  & \textbf{Value} & \textbf{$p$-value} & \textbf{$\hat{A}_{12}$} & \textbf{Value} & \textbf{$p$-value} & \textbf{$\hat{A}_{12}$} & \textbf{Value} & \textbf{$p$-value} & \textbf{$\hat{A}_{12}$} & \textbf{Value} & \textbf{$p$-value} & \textbf{$\hat{A}_{12}$} \\
\cmidrule(lr){1-1}\cmidrule(lr){2-2}\cmidrule(lr){3-3}\cmidrule(lr){4-4}\cmidrule(lr){5-5}\cmidrule(lr){6-6}\cmidrule(lr){7-7}\cmidrule(lr){8-8}\cmidrule(lr){9-9}\cmidrule(lr){10-10}\cmidrule(lr){11-11}\cmidrule(lr){12-12}\cmidrule(lr){13-13}\cmidrule(lr){14-14}
\Full & \ChiSquare & 54.30$\,\pm\,$0.45 & 1.72E-04 & $0.00$ & 52.31$\,\pm\,$0.26 & 1.73E-04 & $0.00$ & 99.87$\,\pm\,$0.06 & 1.58E-04 & $1.00$ & 68.64$\,\pm\,$0.22 & 1.73E-04 & $0.00$ \\
 & \JSDiv & 52.64$\,\pm\,$0.90 & 1.63E-04 & $0.00$ & 51.42$\,\pm\,$0.49 & 1.63E-04 & $0.00$ & 99.89$\,\pm\,$0.04 & 1.16E-04 & $1.00$ & 67.87$\,\pm\,$0.43 & 1.63E-04 & $0.00$ \\
 & \Inverse & 50.01$\,\pm\,$0.01 & 1.03E-04 & $0.00$ & 50.00$\,\pm\,$0.01 & 1.03E-04 & $0.00$ & 100.00$\,\pm\,$0.00 & 5.47E-05 & $1.00$ & 66.67$\,\pm\,$0.01 & 1.03E-04 & $0.00$ \\
 & \DOSS & 65.03$\,\pm\,$0.19 & 1.72E-04 & $0.00$ & 60.41$\,\pm\,$0.16 & 1.73E-04 & $0.00$ & 98.64$\,\pm\,$0.08 & 1.52E-04 & $1.00$ & 74.42$\,\pm\,$0.12 & 1.73E-04 & $0.00$ \\
 & \textbf{\ApproachName} & \textbf{85.65$\,\pm\,$0.09} & -- & -- & \textbf{86.75$\,\pm\,$0.07} & -- & -- & 84.84$\,\pm\,$0.14 & -- & -- & \textbf{85.60$\,\pm\,$0.09} & -- & -- \\
\cmidrule(lr){1-1}\cmidrule(lr){2-2}\cmidrule(lr){3-3}\cmidrule(lr){4-4}\cmidrule(lr){5-5}\cmidrule(lr){6-6}\cmidrule(lr){7-7}\cmidrule(lr){8-8}\cmidrule(lr){9-9}\cmidrule(lr){10-10}\cmidrule(lr){11-11}\cmidrule(lr){12-12}\cmidrule(lr){13-13}\cmidrule(lr){14-14}
\Partial & \QOIN & 61.69$\,\pm\,$0.26 & 1.41E-04 & $0.00$ & 57.68$\,\pm\,$0.21 & 6.34E-05 & $0.00$ & 99.71$\,\pm\,$0.18 & 1.32E-02 & $0.17$ & 72.77$\,\pm\,$0.17 & 1.49E-04 & $0.00$ \\
 & \QOPS* & 49.97$\,\pm\,$0.00 & N/A & N/A & 49.98$\,\pm\,$0.00 & N/A & N/A & 94.07$\,\pm\,$0.00 & N/A & N/A & 65.28$\,\pm\,$0.00 & N/A & N/A \\
 & \textbf{\ApproachName} & \textbf{99.95$\,\pm\,$0.09} & -- & -- & \textbf{100.00$\,\pm\,$0.00} & -- & -- & \textbf{99.90$\,\pm\,$0.19} & -- & -- & \textbf{99.95$\,\pm\,$0.10} & -- & -- \\
\bottomrule[1pt]
\end{tabular}

        }
        {\justify\footnotesize
        We mark * for \QOPS to indicate that it performs only one execution per test case, so the reported metrics show no variance. 
        Statistical tests for \QOPS are denoted as N/A because it was run once, whereas the other approaches were repeated 10 times. For each test, the first group is the listed baseline and the second is \ApproachName; thus, $\hat{A}_{12}<0.5$ indicates lower values for the baseline, and $\hat{A}_{12}>0.5$ indicates higher values.
        \par}
    \end{center}
    
\end{table}

\begin{figure}[htbp]
    \centering
    
    \subfloat[Approaches' accuracy across programs.\label{fig:RQ1_program}]{%
        \includegraphics[width=0.495\linewidth]{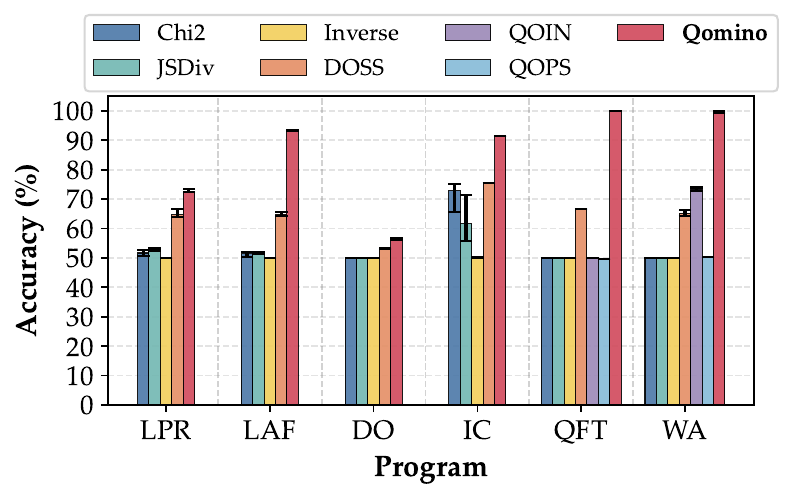}%
    }
    \hfill
    \subfloat[\ApproachName's effectiveness across fake backends.\label{fig:RQ1_backend}]{%
        \includegraphics[width=0.475\linewidth]{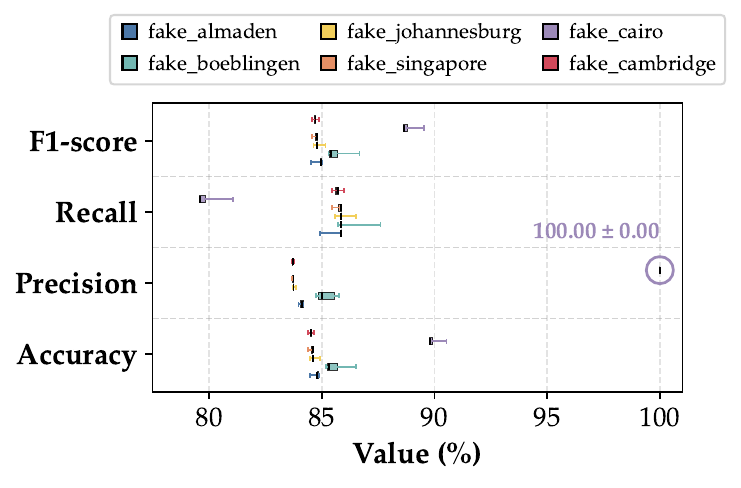}%
    }
    
    \caption{Fine-grained investigation of effectiveness, broken down by program and by backend.}
    \label{fig:RQ1_overall}
    
\end{figure}

Figure~\ref{fig:RQ1_program} gives a program-level view of accuracy. \ApproachName ranks among the strongest approaches on every program, with accuracy generally falling in the upper part of the plotted range (roughly $70\%$--$100\%$) and exceeding 90\% in 4/6 programs. Most baselines, by contrast, stay near the $50\%$ chance level on several programs and rarely exceed it much.
Nevertheless, all approaches perform worse on \DO than on most of the other programs, suggesting that \DO is intrinsically more difficult for noise-aware testing. One plausible contributing factor is that \DO reported by the original paper~\cite{li2026dynamic} has the largest circuit size  
and depth among the six programs (corresponding to 4,179 and 4,117 in Section~\ref{sec:benchmark}). 
Longer noisy execution provides more opportunities for gate and decoherence errors to distort the measured distribution, thereby the potential violation to our assumption that the expected output state should be the dominant one.

Figure~\ref{fig:RQ1_backend} further examines whether \ApproachName's effectiveness is consistent across the six simulated backends. Results of the four metrics of \ApproachName are broadly consistent across \texttt{fake\_almaden}, \texttt{fake\_johannesburg}, \texttt{fake\_singapore}, and \texttt{fake\_cambridge}, where the plotted values remain approximately within the $80\%$--$100\%$ interval.
Such consistency suggests that the dominance assumption underlying \ApproachName holds under different backend noise characteristics being discussed, and therefore that the test results are reliable.
The remaining two backends show visibly different metric profiles, where especially, testing on \texttt{fake\_cairo} even produces nearly 100.00\% precision, 
which indicates that backend-specific noise can still affect the balance between false positives (precision) and false negatives (recall). Nevertheless, the variation does not overturn the overall conclusion from Table~\ref{tab: rq1_overall}: \ApproachName retains high and comparatively balanced effectiveness across the backend set.

\AnswerToRQ{1}{\ApproachName produces the most reliable and balanced test results in our controlled benchmark, achieving $85.65\%$ overall accuracy on all six programs (\Full group) and $99.95\%$ on the \QFT and \WA subset (\Partial group). It significantly outperforms all \Full baselines in accuracy, precision, and F1-score, and \QOIN in all four \Partial-group metrics. Although the \Full baselines attain higher recall, this reflects their strong tendency to classify tests as failures and is accompanied by substantially lower accuracy and precision. \ApproachName's effectiveness is also broadly consistent across the evaluated programs and backends.}

\subsubsection{RQ2: Cost Comparison}\label{sec:cost_comparison}

Table~\ref{tab:rq2_cost_overall} reports the total execution time in $10^4$ seconds of running all programs discussed in corresponding groups on all the six backends. 
As expected, \ChiSquare and \JSDiv are faster, because they only compare the measured distributions and do not introduce additional quantum operations or complicated post-processing.
Nevertheless, \ApproachName is consistently more efficient than the quantum-specific \Inverse and \DOSS approaches.
Even the inference-only cost of \QOIN is already higher than \ApproachName; \QOIN requires additional time for model training and additional memory for model assets.
\QOPS is the clear cost outlier: its total time even reaches $20.30\times10^4$ seconds (56.39 hours), which explains why we can only afford to execute it once in our experiment and only restrict it to the partial evaluation group.
The original studies of \QOPS and \QOIN do not report execution-time measures; future evaluations of noise-aware QST would benefit from reporting execution cost alongside effectiveness.

\begin{table}[htbp]
	\centering
	\begin{minipage}[t]{0.50\linewidth}
		\vspace{0pt}
		\centering
		\caption{Total execution time ($10^4$ s) as the sum of running all selected programs on all backends.}
		\label{tab:rq2_cost_overall}
		\resizebox{\linewidth}{!}{%
\begin{tabular}{c|c|cccc}
\toprule[1pt]
\textbf{Group} & \textbf{Approach} & \multicolumn{4}{c}{\textbf{Total time ($10^4$ s)}} \\
\cmidrule(lr){3-6}
 & & \textbf{Value} & \textbf{$p$-value} & \textbf{$\hat{A}_{12}$} & \textbf{Magnitude} \\
\cmidrule(lr){1-1}\cmidrule(lr){2-2}\cmidrule(lr){3-3}\cmidrule(lr){4-4}\cmidrule(lr){5-5}\cmidrule(lr){6-6}
\Full & \ChiSquare & 1.16$\,\pm\,$0.06 & 1.83E-04 & $0.00$ & Large \\
 & \JSDiv & 1.16$\,\pm\,$0.06 & 1.83E-04 & $0.00$ & Large \\
 & \Inverse & 2.97$\,\pm\,$0.05 & 1.83E-04 & $1.00$ & Large \\
 & \DOSS & 3.06$\,\pm\,$0.14 & 1.83E-04 & $1.00$ & Large \\
 & \textbf{\ApproachName} & 1.66$\,\pm\,$0.06 & -- & -- & -- \\
\cmidrule(lr){1-1}\cmidrule(lr){2-2}\cmidrule(lr){3-3}\cmidrule(lr){4-4}\cmidrule(lr){5-5}\cmidrule(lr){6-6}
\Partial & \QOIN & 3.00$\,\pm\,$0.03 & 1.83E-04 & $1.00$ & Large \\
 & \QOPS* & 20.30$\,\pm\,$0.00 & N/A & N/A & N/A \\
 & \textbf{\ApproachName} & \textbf{0.49$\,\pm\,$0.02} & -- & -- & -- \\
\bottomrule[1pt]
\end{tabular}
		}
	\end{minipage}
    \hfill
	\begin{minipage}[t]{0.46\linewidth}
		\vspace{0pt}
		\centering
		\includegraphics[width=\linewidth]{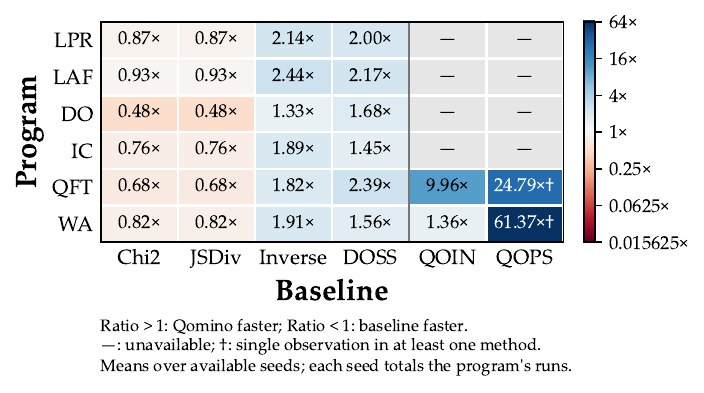}
		\captionof{figure}{Ratio of mean execution time across programs.}
		\label{fig:RQ2_program}
	\end{minipage}
\end{table}

Figure~\ref{fig:RQ2_program} records the ratio (baseline / \ApproachName) of mean execution time over repeated seeds, where the time corresponds to the wall-clock time of running one program on all backends.
The result illustrates that \ApproachName is always faster than \Inverse and \DOSS across all six programs. Compared to other programs, on \DO, its advantage over \Inverse is smallest, while its cost relative to \ChiSquare and \JSDiv is highest. The latter may be related to \DO's larger circuit size and depth: applying inverse operations adds gates to an already large circuit, potentially widening the cost gap from methods that only compare measured distributions without additional quantum operations.

\begin{table}[htbp]

    \caption{Total execution time ($10^3$ s) for testing approaches across backends.}\label{tab: rq2_cost}
    
    \begin{center}
        \resizebox{0.98\columnwidth}{!}{
            \begin{tabular}{c|cccc|c|cc|c}
\toprule[1pt]
\textbf{Backend} & \multicolumn{5}{c|}{\textbf{\Full}} & \multicolumn{3}{c}{\textbf{\Partial}} \\
\cmidrule(lr){2-6}\cmidrule(lr){7-9}
 & \ChiSquare & \JSDiv & \Inverse & \DOSS & \textbf{\ApproachName} & \QOIN & \QOPS* & \textbf{\ApproachName} \\
\cmidrule(lr){1-1}\cmidrule(lr){2-2}\cmidrule(lr){3-3}\cmidrule(lr){4-4}\cmidrule(lr){5-5}\cmidrule(lr){6-6}\cmidrule(lr){7-7}\cmidrule(lr){8-8}\cmidrule(lr){9-9}
\texttt{fake\_almaden} & 1.97$\,\pm\,$0.10 & 1.97$\,\pm\,$0.10 & 4.84$\,\pm\,$0.05 & 4.66$\,\pm\,$0.30 & 2.80$\,\pm\,$0.10 & 4.98$\,\pm\,$0.06 & 33.97$\,\pm\,$0.00 & \textbf{0.82$\,\pm\,$0.05} \\
\texttt{fake\_boeblingen} & 1.79$\,\pm\,$0.08 & 1.79$\,\pm\,$0.08 & 4.54$\,\pm\,$0.08 & 4.07$\,\pm\,$0.19 & 2.59$\,\pm\,$0.08 & 5.22$\,\pm\,$0.03 & 30.85$\,\pm\,$0.00 & \textbf{0.76$\,\pm\,$0.04} \\
\texttt{fake\_johannesburg} & 1.28$\,\pm\,$0.04 & 1.28$\,\pm\,$0.04 & 3.75$\,\pm\,$0.18 & 3.14$\,\pm\,$0.28 & 1.95$\,\pm\,$0.05 & 4.64$\,\pm\,$0.06 & 19.41$\,\pm\,$0.00 & \textbf{0.50$\,\pm\,$0.02} \\
\texttt{fake\_singapore} & 1.88$\,\pm\,$0.09 & 1.88$\,\pm\,$0.09 & 4.75$\,\pm\,$0.08 & 6.02$\,\pm\,$0.24 & 2.70$\,\pm\,$0.10 & 4.90$\,\pm\,$0.07 & 32.25$\,\pm\,$0.00 & \textbf{0.78$\,\pm\,$0.04} \\
\texttt{fake\_cairo} & 1.42$\,\pm\,$0.08 & 1.42$\,\pm\,$0.08 & 4.48$\,\pm\,$0.24 & 5.35$\,\pm\,$0.27 & 2.06$\,\pm\,$0.05 & 4.87$\,\pm\,$0.06 & 26.95$\,\pm\,$0.00 & \textbf{0.70$\,\pm\,$0.01} \\
\texttt{fake\_cambridge} & 3.22$\,\pm\,$0.18 & 3.22$\,\pm\,$0.19 & 7.38$\,\pm\,$0.13 & 7.33$\,\pm\,$0.56 & 4.46$\,\pm\,$0.21 & 5.44$\,\pm\,$0.04 & 59.53$\,\pm\,$0.00 & \textbf{1.36$\,\pm\,$0.10} \\
\bottomrule[1pt]
\end{tabular}

        }
        {\justify\footnotesize
        We mark * for \QOPS to indicate that it performs only one execution per test case, so zero variance does not reflect statistical stability across multiple runs.
        \par}
    \end{center}
    
\end{table}

Table~\ref{tab: rq2_cost} shows that the cost of \ApproachName varies across backends. In the \Full scope, it is lowest on \texttt{fake\_johannesburg} ($1.95\times10^3$ seconds) and highest on \texttt{fake\_cambridge} ($4.46\times10^3$ seconds), while in the \Partial scope, it ranges from $(0.50$--$1.36)\times10^3$ seconds.
This variation arises because transpilation to different backends can produce different quantum circuits, thereby changing the time required to execute the circuits even when the PUTs and test cases remain the same.


\AnswerToRQ{2}{\ApproachName has a moderate execution cost and is substantially faster than the quantum-specific approaches (\Inverse, \DOSS, and especially \QOPS) as well as the learning-based noise-mitigation approach \QOIN. The above adavantage still holds in the respective aspects of programs and backends.}

\subsubsection{RQ3: Ablation Study}
Table~\ref{tab: rq3_ablation} reports the cost and effectiveness of \ApproachName and its four variants when evaluation on \texttt{fake\_almaden}.

As for effectiveness, \ApproachName achieves the highest accuracy ($84.78\%$), with statistically significant improvements over the variants and large effect sizes. Removing decomposition causes the largest drop, to $58.59\%$, because applying one full inverse to the entire circuit adds many gates and increases circuit depth. In a noisy quantum execution, each additional gate gives decoherence and gate errors more opportunities to perturb the state amplitudes before measurement, such that measurement distribution can deviate from the expected bitstring even when the program is correct, making the test oracle more likely to report a wrong test result. 
Under the tested random-search alternative, \NoBipartition achieves nearly the same mean accuracy as \ApproachName ($84.71\%$ vs. $84.78\%$) and similar runtime, i.e., (2.79 vs. 2.80)$\times10^3$ s. Although the accuracy difference is statistically significant ($p\text{-value}=0.0199$), its absolute size is small. This ablation tests decomposition with one alternative search strategy; it does not establish bipartition search as optimal, and broader comparisons, including genetic algorithms, remain future work. 
The much lower accuracy of \NoDominance ($50.00\%$) indicates that requiring every measurement to match the expected output is too strict in the evaluated noisy setting. Removing the Hoeffding-guided decision and retry pipeline (\NoHoeffding) reduces accuracy to $80.47\%$, supporting their combined contribution, where this pipeline retains borderline cases as \Inconclusive and collects additional shots within the retry budget before assigning a final result.

\begin{table}[htbp]

	\caption{Cost and effectiveness evaluation for the ablation study.}\label{tab: rq3_ablation}
	
	\begin{center}
		\resizebox{0.98\columnwidth}{!}{
			\begin{tabular}{c|c|ccc|c|ccc}
\toprule[1pt]
\textbf{Approach} & \multicolumn{4}{c|}{\textbf{Cost}} & \multicolumn{4}{c}{\textbf{Effectiveness}} \\
\cmidrule(lr){2-5} \cmidrule(lr){6-9}
 & \textbf{Total time ($10^3$ s)} & \textbf{$p$-value} & \textbf{$\hat{A}_{12}$} & \textbf{Magnitude} & \textbf{Accuracy (\%)} & \textbf{$p$-value} & \textbf{$\hat{A}_{12}$} & \textbf{Magnitude} \\
\cmidrule(lr){1-1}\cmidrule(lr){2-2}\cmidrule(lr){3-3}\cmidrule(lr){4-4}\cmidrule(lr){5-5}\cmidrule(lr){6-6}\cmidrule(lr){7-7}\cmidrule(lr){8-8}\cmidrule(lr){9-9}
\NoDecomposition & 5.87$\,\pm\,$0.07 & 1.83E-04 & $1.00$ & Large & 58.59$\,\pm\,$0.57 & 1.54E-04 & $0.00$ & Large \\
\NoBipartition & 2.79$\,\pm\,$0.05 & 8.50E-01 & $0.47$ & Negligible & 84.71$\,\pm\,$0.12 & 1.99E-02 & $0.20$ & Large \\
\NoDominance & 2.02$\,\pm\,$0.04 & 1.83E-04 & $0.00$ & Large & 50.00$\,\pm\,$0.00 & 5.35E-05 & $0.00$ & Large \\
\NoHoeffding & 2.02$\,\pm\,$0.04 & 1.83E-04 & $0.00$ & Large & 80.47$\,\pm\,$2.60 & 2.15E-04 & $0.01$ & Large \\
\midrule
\textbf{\ApproachName} & 2.80$\,\pm\,$0.10 & -- & -- & -- & \textbf{84.78$\,\pm\,$0.12} & -- & -- & -- \\
\bottomrule[1pt]
\end{tabular}

		}
		{\justify\footnotesize
			Total time refers to the wall-clock time of executing all test cases of all PUTs.
			Metrics for statistical tests ($p$-value, $\hat{A}_{12}$, and Magnitude) are calculated by comparing the corresponding variant with the original \ApproachName. For $\hat{A}_{12}\in [0, 1]$, a value $<0.5$ indicates that the metric value of one variant is lower than that of \ApproachName, and vice versa.
			\par}
	\end{center}
	
\end{table}

Regarding cost, state decomposition transparently reduces execution time by avoiding the full inverse operation, where \NoDecomposition is roughly twice as expensive as \ApproachName. The cost of \NoBipartition is nearly identical to that of \ApproachName, which may be resulted from a limited search space for the state decomposition. 
Together with the time cost of \NoBipartition, we can infer that the automatic state decomposition of \ApproachName does not strongly depend on one specific search algorithm, and remains generally desirable performance.
Although \NoDominance and \NoHoeffding are faster because they remove, respectively, the dominance criterion and the Hoeffding-guided retry process, their lower accuracy shows that these shortcuts reduce execution cost at the expense of less reliable test decisions, making them still worse than \ApproachName under the priority of reliable test results.

\AnswerToRQ{3}{In the ablation on \texttt{fake\_almaden}, \ApproachName achieves $84.78\%$ accuracy at $2.80\times10^3$ s. Without state decomposition, accuracy falls to $58.59\%$ and runtime roughly doubles; removing dominance reduces accuracy to $50.00\%$. Disabling the Hoeffding-guided decision and retry pipeline also lowers accuracy ($80.47\%$), supporting their combined contribution.}

\subsubsection{RQ4: Retry Investigation}\label{sec:rq4}

\begin{figure}[htbp]
	\centering
	
	\subfloat[Proportion of test cases triggerring the retry phase.\label{fig:RQ4_retry_rate}]{%
		\includegraphics[width=0.48\linewidth]{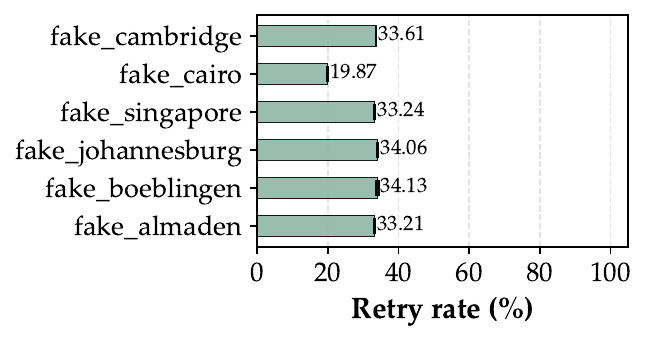}%
	}
	\hfill
	\subfloat[Accuracy reduction of \InconclusiveToPass over \ApproachName.\label{fig:RQ4_incon2pass}]{%
		\includegraphics[width=0.48\linewidth]{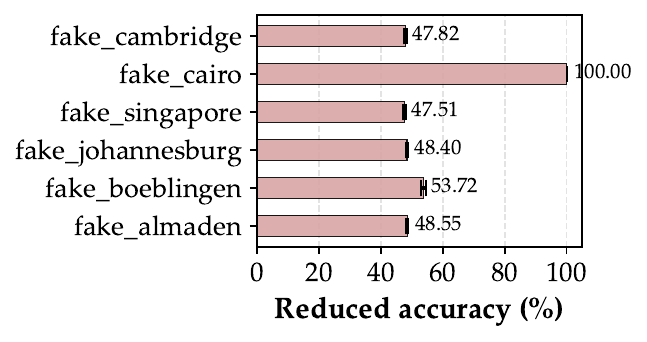}%
	}
	\caption{Investigation of the retry mechanism across the backends.}
	\Description{Two side-by-side plots showing the retry rate on the left and the accuracy reduction with the incon2pass variant on the right, across six noisy backends.}
	
	\label{fig:RQ4_retry}
\end{figure}

Figure~\ref{fig:RQ4_retry_rate} reports the fraction of test cases for which the initial Hoeffding-based decision is \Inconclusive and therefore requires another execution with more shots through the retry phase. The retry rate ranges from $19.87\%$ on \texttt{fake\_cairo} to $34.13\%$ on \texttt{fake\_boeblingen}, with the other backends close to one third of all tests. Thus, with the default shot budget, inconclusive decisions occur in roughly one-fifth to one-third of tests across the evaluated noisy backends, making retries a recurring rather than exceptional outcome in our benchmark.
Moreover, the variation across backends indicates that backend-specific noise characteristics affect how often the available evidence is insufficient for a confident output. Retrying gives the test oracle an opportunity to collect additional measurements before committing to pass or fail, which is, more generally, essential for avoiding premature decisions in noise-aware QST.

To isolate the effect of the final output identification, Figure~\ref{fig:RQ4_incon2pass} compares \ApproachName with \InconclusiveToPass (Section~\ref{sec:baselines}). The plotted metric is the accuracy of \ApproachName minus that of \InconclusiveToPass merely for tests that trigger the retry phase; this metric therefore measures the penalty from replacing \ApproachName's strict treatment of an exhausted \Inconclusive decision (i.e., reporting \Fail) with a direct \Pass.
The reduction is substantial on every backend, ranging from $47.51\%$ on \texttt{fake\_singapore} to $100.00\%$ on \texttt{fake\_cairo}. 
The reason could be that: under the Hoeffding-based decision, satisfying the dominance criterion defined by \ApproachName provides sufficient evidence for passing, whereas failing to satisfy it does not necessarily imply a failure. Directly converting an exhausted \Inconclusive result to \Pass therefore treats tests without sufficient dominance evidence as passing, producing false negatives when genuinely failing tests remain \Inconclusive.

\AnswerToRQ{4}{Retries are triggered for roughly one fifth to one third of tests across backends, indicating that inconclusive decisions are common under noise. The conservative fallback of \ApproachName substantially outperforms \InconclusiveToPass, whose accuracy drops by $47.51\% $--$100.00\%$ on retry-triggering tests.}



\section{Threats to Validity}\label{sec:threats}
\textbf{Internal validity.}
Differences in implementation and execution may affect the comparison. We independently implement the baselines and ablation variants, while deriving the \InconclusiveToPass analysis variant from \ApproachName's runtime log because it changes only the final fallback decision. We use the published parameters or artifact defaults for existing approaches and keep the user-configurable initial shot, decision thresholds, retry budget, and execution environment consistent across approaches and variants.
The remaining asymmetry is that \ApproachName may consume additional shots when the initial execution is \Inconclusive, whereas some approaches use only the initial budget $s_0$. The fixed retry budget ($b_{\text{retry}}=1$) bounds this additional consumption. In our experiments, \ApproachName remains within a reasonable execution time, and the small standard deviations indicate limited variability in cost and effectiveness.

\textbf{Construct validity.}
We measure effectiveness as agreement between test outcomes and the benchmark ground truth using accuracy, precision, recall, and F1-score, rather than counting reported faults. This choice mitigates the risk that measurement- or noise-induced false reports are interpreted as detected program faults.
Besides, time cost is considered for the cost-effectiveness trade-off, which helps to assess the practical applicability of QST approaches under noise.

\textbf{External validity.}
Our discussion of \ApproachName's methodology assumes Pauli-Z measurements because they are implemented as the default measurement basis in many quantum SDKs and are used by the majority of existing QST studies~\cite{li2026methodological}. Still, \ApproachName can be adapted to other measurement bases in theory, by applying the corresponding basis transformation to the specification state.
Although we reuse real-world scalable subject programs and controlled test suites, where the benchmark still contains 1,500 distinct quantum circuits from 30 buggy variants, the results may not always generalize to other programs or mutants. 
Considering high execution cost and limited backend accessibility, we follow the common practice of evaluating QST with classical noisy simulation~\cite{li2026methodological}. Real devices may exhibit noise correlations, calibration drift, and hardware-specific effects that are not captured by these simulators. As advances in quantum hardware, calibration, and noise-suppression techniques reduce the noise affecting quantum computations~\cite{shor1996fault,campbell2017roads}, the expected output observed in the software layer is more likely to remain dominant. Moreover, the configurable dominance strength $\gamma$ can be adjusted to accommodate different noise levels, potentially broadening the range of settings in which \ApproachName applies. 
Finally, our study compares multiple approaches across subject programs, noisy backends, and random seeds, providing a broad evaluation while keeping the comparison tractable. We use fixed, reasonable settings for $s_0$ (initial shot budget), $\delta$ (error probability), $b_{\text{retry}}$ (retry budget), and $\gamma$ (dominance strength) across this experimental matrix so that differences in effectiveness and cost reflect the approaches themselves rather than additional parameter tuning. This design supports a controlled and fair comparison; a broader sensitivity analysis of these parameters remains future work.

\textbf{Conclusion validity.}
We narrow our conclusion to the discussed experimental configurations without making overly broad claims about generality.
Random seeds and backend-dependent transpilation can affect effectiveness and execution time. We repeat affordable runs with 10 random seeds, report means with sample standard deviations, and perform statistical tests, where \QOPS is the exception because repeated execution is prohibitively costly. We also examine program- and backend-level results and analyze retry-triggering tests separately in RQ4, rather than relying only on aggregate metrics. These practices strengthen the comparisons, while the conclusions remain bounded by the tested benchmark and noise models.

\section{Related Work}\label{sec:related}
According to the state of NISQ devices, QSE research has gradually shifted from purely ideal simulation to noise-aware execution~\cite{murillo2025quantumsoftwareengineering, leite2025testing, paltenghi2024survey}.
Recent studies have examined how noise affects QSE activities, focusing on modeling or characterizing quantum noise~\cite{muqeet2024approximating, maschek2025make, bensoussan2025toward}, distinguishing software bugs from hardware noise~\cite{virani2025distinguishing}, and mitigating quantum noise~\cite{muqeet2024machine, muqeet2024mitigatingno, ma2026reliableoracleverification2}. 
These efforts highlight the importance of accounting for quantum noise in QSE; accordingly, our work incorporates noise-induced output variation into test-oracle analysis and quantifies the statistical uncertainty in estimating output dominance from a finite number of shots.

QST research has gained increasing attention in recent years.
Early QST research focused primarily on systematic test generation and adequacy for quantum circuits. \Quito defined input, output, and input--output coverage criteria with statistical test oracles~\cite{ali2021quito}; \QuSBT applied search-based optimization to generate test suites likely to expose failures~\cite{wang2022qusbt}; and \QuCAT used combinatorial testing to cover interactions among program inputs~\cite{wang2023qucat}. These approaches laid important foundations for automated QST, but do not handle backend noise.

More recent QST studies have extended testing beyond fixed-size circuits to scalable quantum programs and have used formula-based specifications to assess program functionality~\cite{li2025preparation,li2026dynamic,li2026benchmarking,long2024testingmultisubroutine,long2025blackbox}. Classical simulation remains common in comprehensive empirical studies, while the impact of noise on QST has received increasing attention~\cite{li2026methodological}. Studies using noisy backends have reported non-negligible effects on test outcomes~\cite{sato2025bug,shao2026quanforge,klymenko2026context,li2026dynamic,acharya2021test}. However, in most of these studies, noise serves primarily as an evaluation condition rather than a factor handled by the testing method.

Noise-focused QST approaches have incorporated zero-noise extrapolation~\cite{muqeet2024quantum} and machine-learning-based noise mitigation~\cite{ma2026reliableoracleverification2,pontolillo2025ideal,muqeet2024mitigatingno}. In contrast, \ApproachName analyzes noisy observations without a separate noise estimation or mitigation phase. Our comparison with \QOIN suggests that learned mitigation may generalize poorly when models are specific to a program--backend pair, and its training and inference costs may be disproportionate for relatively simple circuits.

Besides, as introduced in Section~\ref{sec:methodology}, although prior studies have respectively investigated state decomposition~\cite{li2026dynamic}, inverse operations~\cite{long2024testingmultisubroutine,miranskyy2025feasibility}, and retry mechanisms~\cite{luo2026qemi}, \ApproachName contributes a coordinated workflow in which each idea is substantially improved and adapted to address a distinct cost or reliability bottleneck in noisy QST. Automatic state decomposition reduces the overhead and potential noise amplification of inverse construction and bounded retries resolve inconclusive decisions in our dominance-based criterion. These mechanisms are integrated into a single automated QST pipeline rather than applied independently.

\section{Conclusion and Future Work}\label{sec:conclusion}
In the noisy intermediate-scale quantum (NISQ) era, quantum software testing (QST) should distinguish program failures from deviations caused by measurement uncertainty and backend noise. 
To address this problem, we presented a novel testing approach \ApproachName tailored to noisy backends, which serially decomposes the expected output state, evaluates output dominance with a Hoeffding-based decision, and retries inconclusive tests within a bounded shot budget. 
Our empirical study demonstrates that \ApproachName produces more reliable test results than the compared six baselines, while maintaining a moderate execution cost. The ablation results support the contributions of state decomposition, dominance evaluation, and the retry-aware decision pipeline, and the retry study shows that additional executions are useful for resolving statistically inconclusive outcomes. These support \ApproachName as a practical approach for improving the reliability of QST under noise.

Our evaluation is conducted in a controlled simulation setting with fixed parameters, keeping the multi-factor comparison feasible and fair across approaches.
Future work can explore larger circuits and more quantum programs, mutation patterns, and noisy backends, as well as comprehensively investigate the hyperparameter sensitivity.
Another direction is to adapt the dominance strength $\gamma$ to circuit characteristics, improving reliability--cost tradeoffs across heterogeneous testing tasks.

\section*{Data Availability}\label{sec:data}
We make artifacts of \ApproachName approach and experiments publicly available at~\cite{qolumbina_tester} and~\cite{qomino_experiments}, respectively.
We encapsulate \ApproachName in a callable interface used throughout the experiments to support its reuse in future studies.

\bibliographystyle{plainnat}
\bibliography{sample-base}

\end{document}